\documentstyle[preprint,tighten,aps]{revtex}
\begin{document}

\def\beqar{\begin{eqnarray}}
\def\eeqar{\end{eqnarray}}
\def\be{\begin{eqnarray}}
\def\ee{\end{eqnarray}}
\def\beqast{\begin{eqnarray*}}
\def\eeqast{\end{eqnarray*}}
\def\be{\begin{enumerate}}
\def\ee{\end{enumerate}}
\def\lag{\langle}
\def\rag{\rangle}
\def\fnote#1#2{\begingroup\def\thefootnote{#1}\footnote{#2}
\addtocounter{footnote}{-1}\endgroup}
\def\beq{\begin{equation}}
\def\eeq{\end{equation}}
\def\haf{\frac{1}{2}}
\def\pa{\partial}
\def\plb#1#2#3#4{#1, Phys. Lett. {\bf #2B}, #3 (#4)}
\def\npb#1#2#3#4{#1, Nucl. Phys. {\bf B#2}, #3 (#4)}
\def\prd#1#2#3#4{#1, Phys. Rev. {\bf D#2}, #3 (#4)}
\def\prl#1#2#3#4{#1, Phys. Rev. Lett. {\bf #2}, #3 (#4)}
\def\mpl#1#2#3#4{#1, Mod. Phys. Lett. {\bf A#2}, #3 (#4)}
\def\rep#1#2#3#4{#1, Phys. Rep. {\bf #2}, #3 (#4)}
\def\llp#1#2{\lambda_{#1}\lambda'_{#2}}
\def\lplp#1#2{\lambda'_{#1}\lambda'_{#2}}
\def\slash#1{#1\!\!\!\!\!/}


\draft
\preprint{
\begin{tabular}{r}
KAIST-TH 97/02
\\
hep-ph/9701283
\end{tabular}
}
\title{
Constraints on the R-parity- and Lepton-Flavor-Violating
Couplings from $B^0$ Decays to Two Charged Leptons.
}
\author{
Ji-Ho Jang 
\thanks{Electronic address: jhjang@chep6.kaist.ac.kr},
Jae Kwan Kim,
and
Jae Sik Lee
\thanks{Electronic address: jslee@chep6.kaist.ac.kr}
}
\address{
Department of Physics, Korea Advanced Institute of Science and
Technology,\\
Taejon 305-701, Korea \\
}

\maketitle

\begin{abstract}
We derive the upper bounds on certain products of R-parity- and
lepton-flavor-violating couplings from the decays of the neutral
$B$ meson into two charged leptons. These modes of $B^0$ decays
can constrain the product combinations of the couplings with
one or more heavy generation indices.
We find that most of these bounds are
stronger than the previous ones.
\end{abstract}


In the standard model (SM), there are no couplings which violate
baryon number $B$ and lepton number $L$. 
The failures of experimental searches to find $B$-
and/or $L$-violating processes show that this feature
of the SM is a good one. But one notes that this feature of the SM 
is not the result of gauge invariance. 
In the supersymmetric standard models, there are
gauge invariant interactions which violate $B$ and $L$
generally. To prevent occurrences of these $B$- and $L$-violating
interactions in the supersymmetric standard models, 
an additional global symmetry is required.
This requirement leads to the consideration of
the so-called R-parity.
$R$ parity is given by the relation $R_p=(-1)^{(3B+L+2S)}$ where 
$S$ is the intrinsic spin of a field. 
According to this definition, $R_p$ of the ordinary SM particles
is $+1$ and $R_p$ of the superpartners is $-1$.
Even though the requirement of $R_p$ conservation gives a theory
consistent with
present experimental searches, there is no good theoretical justification
for this requirement. Therefore,
the models with explicit $R_p$ violation have been considered by
many authors \cite{ago}. 
If we discover a sign of $R_p$ violations in future experiments, it
may provide us with some hints of the existence of supersymmetry.

In the model without $R_p$, the supersymmetric
particles can decay into the ordinary particles alone. 
So the couplings which violate $R_p$ can be detected by
using usual particle detectors.
To discover the $R_p$ violation in future experiments,
we need to know what kinds of
couplings are severely constrained by present experimental data.
Therefore, it is important to constrain the $R_p$-violating
couplings from the present data, especially data on the processes forbidden 
or highly suppressed in the SM.
Usually, the bounds on the $R_p$-violating couplings 
with heavy fields 
are not stronger than those with at most one heavy field.

In this paper, we try to derive 
the upper bounds on certain products of $R_p$- and
lepton-flavor-violating couplings from the decays of the neutral
$B$ meson into two charged leptons in the minimal supersymmetric
standard model (MSSM) with explicit $R_p$ violation. 
These modes of $B^0$ decays
can constrain the product combinations of the couplings with
one or more heavy generation indices.
We find that most of these bounds are
stronger than the previous ones.

In the MSSM,
the most general $R_p$-violating superpotential is given by
\beq
W_{R\!\!\!\!/_p}=\lambda_{ijk}L_iL_jE_k^c+\lambda_{ijk}'L_iQ_jD_k^c+
\lambda_{ijk}''U_i^cD_j^cD_k^c.
\eeq
Here $i,j,k$ are generation indices and we assume that possible bilinear terms
$\mu_i L_i H_2$ can be rotated away.
$L_i$ and $Q_i$ are the SU(2)-doublet lepton and quark superfields and
$E_i^c,U_i^c,D_i^c$ are the singlet superfields, respectively. 
$\lambda_{ijk}$ and
$\lambda_{ijk}''$ are antisymmetric under the interchange of the first two and
the last two generation indices,
respectively; $\lambda_{ijk}=-\lambda_{jik}$ and
$\lambda_{ijk}''=-\lambda_{ikj}''$. So the number of couplings is 45 (9 of the
$\lambda$ type, 27 of the $\lambda'$ type and 9 of the $\lambda''$ type).
Among these 45 couplings, 36 couplings are related with the lepton
flavor violation.

The simultaneous presence of the $B$- and $L$-violating couplings leads to the
squark-mediated proton decay. To avoid a too rapid proton decay we should
constrain these couplings very strongly: 
\beq
\left| \lambda'\cdot\lambda''\right| \leq 10^{-24}
\eeq
for squark masses around 1 TeV \cite{hinkae}.
Note that this bound does not affect certain products of couplings with
heavy generations at the tree level. 
Therefore, one can expect the possibility of
large $R_p$-violating couplings with heavy generation indices. But
recently, the authors of Ref. \cite{smivi} studied the one-loop structure.
They showed that, choosing whichever
pair of couplings $\lambda'$ and $\lambda''$, there is always at least
one diagram relevant for the proton decay at one-loop level. This means
there are upper bounds on {\it all} products of $\lambda'$ and $\lambda''$ from
the proton decay. The less-suppressed pair of couplings has the bound,
\beq
\left| \lambda'\cdot\lambda''\right| \leq 10^{-9}.
\eeq
It is known that the bounds from the proton decay are better than those 
from $B$-meson decays unless branching ratios of
$10^{-8}$ or better are obtained \cite{thopocarl}.
To constrain $R_p$-violating couplings from the 
$B$-meson decays, therefore, we should avoid the simultaneous presence of
$B$- and $L$-violating couplings.

In this paper we assume that $B$-violating couplings of the $\lambda''$ type
are vanishing.  For example,
one can construct a grand unified model which has only lepton number
nonconserving trilinear operators in the low-energy superpotential when
$R_p$ is broken only by bilinear terms of the form $L_i H_2$
\cite{hasu}.
We observe that many additional and stronger upper bounds on 
$\lambda\lambda'$ and $\lambda'\lambda'$ involving heavy generations
can be derived from the experimental upper bounds on
the pure leptonic decays of the $B$ meson;
$B^0 \rightarrow e^+e^-$ or 
$e^{\pm}\mu^{\mp}$ or
$e^{\pm}\tau^{\mp}$ or
$\mu^{\pm}\tau^{\mp}$ or
$\mu^+\mu^-$.

The exchange of the sleptons and squarks leads to the four-fermion
interactions in the effective Lagrangian at the scale of $B$-meson mass.
Among these four-fermion operators, there is
a term relevant for $B^0$ decays into two charged leptons.
This effective term has 2 down-type quarks and 2 charged leptons.
From Eq. (1) we obtain
\beqar
{\cal L}^{\rm eff}_{2d-2l}&=& 
\sum_{n=1}^{3}\frac{2}{m_{\tilde{\l}_n}^2}\left[\lambda^*_{njk}\lambda'_{nlm}
(\bar{e}_jP_Re_k)(\bar{d}_mP_Ld_l)
+{\rm H.c.}\right]-
\nonumber \\
&& \sum_{n,r,s=1}^{3}\frac{1}{2m_{\tilde{Q}_n}^2}K_{nr}K^*_{ns}
\lambda'^*_{jrk}\lambda'_{lsm}
(\bar{e}_j\gamma^{\mu}P_Le_l)(\bar{d}_m\gamma_{\mu}P_Rd_k),
\eeqar
where $K$ is the Cabibbo-Kobayashi-Maskawa (CKM) matrix and we assume 
the matrices of the soft mass terms are diagonal.
The effective term ${\cal L}^{\rm eff}_{2d-2l}$ is also related with
the pure leptonic decay modes of the neutral $K$ meson \cite{choroy}.
From these decays of $K^0$ meson one can derive upper bounds on
$\lambda\lambda'$ between $10^{-7}$ and $10^{-8}$. 
But for some $\lambda'$, which include at least one heavy generation index,
the product combinations of the $\lambda\lambda'$-type are not
constrained from $K$-meson decays. 
$B$-meson decays can constrain certain product combinations
of $\lambda\lambda'$-type which are not constrained from $K$-meson decays. 
But,
the upper bounds from $B$ decays are less stringent than those from
$K$ decays since the experimental upper bounds on $B$-meson decays
are weaker than those on $K$-meson decays. 

There are upper bounds on {\it single} $R_p$-violating coupling from
several different sources \cite{han,beta,numass,agagra}. 
Among these, upper bounds from
neutrinoless double $\beta$ decay \cite{beta}, $\nu$ mass \cite{numass} and
$K^+,t-$quark decays \cite{agagra} are strong. 
Neutrinoless double $\beta$ decay gives
$\lambda'_{111}<4\times10^{-4}$.  
The bounds from $\nu$ mass are $\lambda_{133}<10^{-3}$ 
and $\lambda'_{133}<10^{-3}$.
From $K^+$-meson decays one obtains $\lambda'_{ijk}<0.012$ for $j=1$ and 2.
Here all masses of scalar partners which mediate the processes are assumed
to be 100 GeV.
Any cosmological bounds can be avoided by assuming the smallest lepton-
number-violating coupling $\lambda'$ is less than $10^{-7}$ \cite{dross}.
Fully reviewed and updated limits on single $R_p$-violating 
coupling can be found in \cite{bha,chahu}.

The measurements of the branching ratios of the $B^0$-meson decays to two
charged leptons give the upper bounds (at 90$\%$ C.L.) \cite{CLEO}
\beqar 
{\cal B}(B_d \rightarrow e^+e^-) &<& 5.9 \times 10^{-6}, \nonumber \\
{\cal B}(B_d \rightarrow e^{\pm}\mu^{\mp}) &<& 5.9 \times 10^{-6}, \nonumber \\
{\cal B}(B_d \rightarrow e^{\pm}\tau^{\mp}) &<& 5.3 \times 10^{-4}, \nonumber \\
{\cal B}(B_d \rightarrow \mu^{\pm}\tau^{\mp}) &<& 8.3 \times 10^{-4},
\eeqar
and \cite{CDF} 
\beqar 
{\cal B}(B_d \rightarrow \mu^+\mu^-) &<& 1.6 \times 10^{-6}, \nonumber \\
{\cal B}(B_s \rightarrow \mu^+\mu^-) &<& 8.4 \times 10^{-6}.
\eeqar

In the SM, the processes which
have different lepton species as decay products are forbidden
due to the conservation of each lepton-flavor number.
On the other hand, the processes with the same lepton species
are highly suppressed;
${\cal B}(B_s \rightarrow e^+ e^-) \simeq (8.0 \pm 3.5) \times 10^{-14}$,
${\cal B}(B_s \rightarrow \mu^+ \mu^-) \simeq (3.5 \pm 1.0) \times 10^{-9}$
\cite{ali}.
The pure leptonic decays of $B_d$ meson are more suppressed by the smaller CKM
angle. So we neglect the SM contributions to the processes under
considerations.
But in the MSSM under considerations, all
processes are possible through ${\cal L}^{\rm eff}_{2d-2l}$ at the tree-level.

Using the PCAC (partial conservation of axial-vector current) relations
\beqar
\lag0|\bar{b}\gamma^{\mu}\gamma_5q|B_q(p)\rag &=&  i f_{B_q} p^{\mu}_{B_q},
\nonumber \\
\lag0|\bar{b}\gamma_5q|B_q(p)\rag &=& - i f_{B_q} \frac{M_{B_q}^2}{m_b+m_q}
\cong - i f_{B_q} M_{B_q},
\eeqar
the decay rate of the
processes $B_q \rightarrow e_i^-e_j^+$ reads
\beqar
\Gamma(B_q && \rightarrow e_i^-e_j^+)=
\frac{f_{B_q}^2 M_{B_q}^3}{64\pi \tilde{m}^4}
\sqrt{1+(\kappa_i^2-\kappa_j^2)^2-2(\kappa_i^2+\kappa_j^2)}
\times \nonumber \\
&&
\left\{ (|{\cal A}_{ij}^q|^2+|{\cal B}_{ij}^q|^2)(1-\kappa_i^2-\kappa_j^2)+
|{\cal C}_{ij}^q|^2\left[(\kappa_i^2+\kappa_j^2)-(\kappa_i^2-\kappa_j^2)^2\right]+
 \right.
\nonumber \\
&&
\left.
4{\rm Re}({\cal A}_{ij}^q{\cal B}_{ij}^{q*})\kappa_i\kappa_j
+2\kappa_j{\rm Re}({\cal A}_{ij}^q{\cal C}_{ij}^{q*})(1+\kappa_i^2-\kappa_j^2)
+2\kappa_i{\rm Re}({\cal B}_{ij}^q{\cal C}_{ij}^{q*})(1+\kappa_j^2-\kappa_i^2)
  \right\},
\eeqar
where $q=1(B_d)$ or $2(B_s)$, $\kappa_i \equiv m_{e_i}/M_{B_q}$.
We assume the universal soft mass $\tilde{m}$.
$M_{B_q}$ is the mass of $B_q$ meson and
$f_{B_q}$ is the usual leptonic decay constant of the $B_q$ meson.
The constants ${\cal A}_{ij}^q,{\cal B}_{ij}^q$, and ${\cal C}_{ij}^q$ 
which depend on the generations of leptons and the type
of decaying neutral $B$ meson are given by
\beqar
{\cal A}_{ij}^q&=& 2\sum_{n=1}^{3}\lambda^{*}_{nij}\lambda'_{nq3}, \nonumber \\
{\cal B}_{ij}^q&=& 2\sum_{n=1}^{3}\lambda_{nji}\lambda'^{*}_{n3q}, \nonumber \\
{\cal C}_{ij}^q&=&\frac{1}{2}\sum_{n,p,s=1}^{3}K_{np}K_{ns}^*
\lambda'^{*}_{ipq}\lambda'_{js3} =
\frac{1}{2}\sum_{n=1}^{3}
\lambda'^{*}_{inq}\lambda'_{jn3} .
\eeqar
Note that 
${\cal A}_{ij}^q \neq {\cal A}_{ji}^q$, etc.
and we assume the universal soft mass.
The values of $\kappa_i$ are the followings: 
$\kappa_1 = 10^{-4}$,
$\kappa_2 = 2\times10^{-2}$, and
$\kappa_3 = 0.34$.
So we neglect the effects of lepton masses
if there is no $\tau$ in the decay products. 

Neglecting lepton masses, the decay rate becomes simple and has
terms which depend only on ${\cal A}_{ij}^q$ and ${\cal B}_{ij}^q$.
Numerically, we obtain
\beqar
\Gamma(B_q && \rightarrow e_i^-e_j^+) \approx
2.93\times10^{-10}\left[|{\cal A}_{ij}^q|^2+|{\cal B}_{ij}^q|^2\right]
\times \nonumber \\
&&
\left(\frac{f_{B_q}}{0.2 ~{\rm GeV}}\right)^2
\left(\frac{M_{B_q}}{5.28 ~{\rm GeV}}\right)^3
\left(\frac{100~ {\rm GeV}}{\tilde{m}}\right)^4
~~{\rm GeV}.
\eeqar
For example, let us think about the decay of $B_d$ into $e^+e^-$.
In this case, $i=1$, $j=1$ and $q=1$. 
Combining the above equation with Eq. (5), we obtain
\beq
|{\cal A}_{11}^1|^2+|{\cal B}_{11}^1|^2 < 8.5 \times 10^{-9}.
\eeq
From Eq. (9),
${\cal A}_{11}^1=2(-\lambda^*_{121}\lambda'_{213}-
\lambda^*_{131}\lambda'_{313})$
and
${\cal B}_{11}^1=2(-\lambda_{121}\lambda'^*_{231}-
\lambda_{131}\lambda'^*_{331})$.
Under the assumption that only one product combination is not zero,
we obtain the same upper bound $4.6 \times 10^{-5}$ on the magnitudes
of four coupling products; 
$\lambda_{121}\lambda'_{213}$,
$\lambda_{121}\lambda'_{231}$,
$\lambda_{131}\lambda'_{313}$, and
$\lambda_{131}\lambda'_{331}$. 
In a similar way, other upper bounds can be obtained
in the cases of $B_d$ decays into $e^-\mu^+ + e^+\mu^-$,
$\mu^+\mu^-$ and $B_s$ decay into $\mu^+\mu^-$, see Table \ref{haha}.

We have two decay modes of $B$ meson which have $\tau$ in the decay products;
$B^0 \rightarrow e^{\pm}\tau^{\mp}$ or
$\mu^{\pm}\tau^{\mp}$. In this case,
\beqar
\Gamma(B_q && \rightarrow \tau^-e_i^+) + 
\Gamma(B_q  \rightarrow e_i^-\tau^+) \approx 2.93\times10^{-10} \times 0.88
\times \nonumber \\
&&\left\{
0.88\times\left(|{\cal A}_{3i}^q|^2+|{\cal B}_{3i}^q|^2+
|{\cal A}_{i3}^q|^2+|{\cal B}_{i3}^q|^2\right) +
0.10\times\left(|{\cal C}_{3i}^q|^2+ |{\cal C}_{i3}^q|^2\right)
\right.+
\nonumber \\
&&
\left.
0.60\times  Re\left({\cal B}_{3i}^q{\cal C}^{q*}_{3i}+
{\cal A}_{i3}^q{\cal C}^{q*}_{i3}\right)
\right\}\times \nonumber \\
&&\left(\frac{f_{B_q}}{0.2 ~{\rm GeV}}\right)^2
\left(\frac{M_{B_q}}{5.28 ~{\rm GeV}}\right)^3
\left(\frac{100~ {\rm GeV}}{\tilde{m}}\right)^4
~~{\rm GeV}.
\eeqar
One obtains the most stringent bounds on
$|{\cal A}_{i3}^q|^2$ and $|{\cal B}_{i3}^q|^2$ or 
on product combinations of the $\lambda\lambda'$-type.
The bounds on
${\cal B}_{3i}^q{\cal C}^{q*}_{3i}$ and
${\cal A}_{i3}^q{\cal C}^{q*}_{i3}$ are slightly weaker than those on
$|{\cal A}_{i3}^q|^2$ and $|{\cal B}_{i3}^q|^2$.
These constrain the product combinations of the type of
$\lambda\lambda'\lambda'\lambda'$.
We will not consider these kinds of bounds on
$\lambda\lambda'\lambda'\lambda'$ since these bounds
seem to be less important.
The bounds on $|{\cal C}_{i3}^q|^2$ and
$|{\cal C}_{3i}^q|^2$ are the weakest ones.
For the case of $B_d$ decay into electron and $\tau$,
$i=1$ and $q=1$. Neglecting the terms of
${\rm Re}[({\cal A,B})_{3i}^q{\cal C}^{q*}_{3i}]$, 
\beq
0.88\times\left(|{\cal A}_{31}^1|^2+|{\cal B}_{31}^1|^2+
|{\cal A}_{13}^1|^2+|{\cal B}_{13}^1|^2\right) +
0.10\times\left(|{\cal C}_{31}^1|^2+ |{\cal C}_{13}^1|^2\right)
< 8.5 \times 10^{-7}.
\eeq
Under the assumption that only one product combination is not zero,
we obtain the same upper bound $4.9 \times 10^{-4}$ on the magnitudes
of eight coupling products of the $\lambda\lambda'$-type.
We also obtain the same upper bound $5.8 \times 10^{-3}$ on the magnitudes
of six coupling products of the $\lambda'\lambda'$-type;
$\lplp{311}{113}$, $\lplp{321}{123}$, $\lplp{331}{133}$, $\lplp{111}{313}$,
$\lplp{121}{323}$, and $\lplp{131}{333}$. Among these, only the bound
on $\lplp{131}{333}$ is stronger than previous one.
In a similar way, the bounds 
coming from $B_d$ decay into $\mu^-\tau^+ + \mu^+\tau^-$
can be obtained, see Table \ref{haha}.
In the case of $B_d$ decay into $\mu$ and $\tau$, 
only two bounds on $\lplp{231}{333}$ and $\lplp{233}{331}$ are stronger
than previous ones.


The previous bounds are calculated from the bounds
on single $R_p$-violating coupling, see Table I
of Ref. \cite{chahu}\footnote{We use $\lambda'_{11k}<0.012$
and $\lambda'_{132}<0.4$ from \cite{agagra} instead of the values
shown in Table I of Ref. \cite{chahu}.}.
We observe that the bounds on the product combinations of 
the $\lambda\lambda'$-type,
which are between $10^{-4}$ and $10^{-5}$,
are stronger than the previous bounds or comparable
except those on $\lambda_{133}\lambda'_{313}$,
$\lambda_{133}\lambda'_{331}$.
This is because there exists a strong upper bound ($10^{-3}$) on
$\lambda_{133}$ from $\nu$ mass. We find that 
only 3 of 12 bounds on the product combinations 
of the $\lambda'\lambda'$-type are stronger than previous ones.

There are bounds on certain product combinations
of the $\lambda\lambda'$-type from the
$K$-meson decays into pure lepton pairs;
$K_L \rightarrow e^+e^-, K_L \rightarrow \mu^+\mu^-$, and 
$K_L \rightarrow e^+\mu^- + e^-\mu^+$ \cite{choroy}. The bounds from
$K$-meson decays are more stringent than those from $B$-meson decays. 
They are between
$10^{-7}$ and $10^{-8}$. We observe that even though the bounds
on $\lambda\lambda'$ from $B$-meson decays are weaker than those from
$K$-meson decays, $B$-meson decays can
constrain the product combinations of the $\lambda\lambda'$-type
which $K$-meson decays cannot constrain. 

To conclude, we have derived the upper bounds on certain products of $R_p$-
and lepton-flavor-violating couplings from the decays of the neutral
$B$ meson into two charged leptons. These modes of $B^0$ decays
can constrain the product combinations of the couplings with
one or more heavy generation indices which the similar decay modes
of $K$ meson cannot constrain.
We find that the most of the bounds on products of 
the $\lambda\lambda'$-type are stronger than the previous ones.
For the bounds on products of the $\lambda'\lambda'$-type, we find
three stronger bounds than previous ones.

\section*{acknowledgements}
Two of us (J.H.J. and J.S.L.) thank K. Choi for his helpful remarks.
This work was supported in part by KAIST Basic Science Research Program
(J.S.L.).

%
%

\begin{table}
\caption{\label{haha} 
Upper bounds on the magnitudes of products of couplings
derived from $B^0$ decays into two charged leptons. 
}
\begin{tabular}{llll}
Decay Mode & Combinations Constrained & Upper bound & Previous bound \\
\hline
$B_d \rightarrow e^+e^-$& 
$\llp{121}{213}$ & 4.6$\times 10^{-5}$ & 4.8 $\times 10^{-4}$
\\
&$\llp{121}{231}$ &4.6$\times10^{-5}$  & 8.8 $\times 10^{-3}$
\\
&$\llp{131}{313}$ &4.6$\times10^{-5}$  & 1.2 $\times 10^{-3}$
\\
&$\llp{131}{331}$ &4.6$\times10^{-5}$  & 2.6 $\times 10^{-2}$
\\
\hline
$B_d \rightarrow e^+\mu^- + e^- \mu^+$& 
$\llp{121}{113}$ & 4.6$\times 10^{-5}$ & 4.8 $\times 10^{-4}$
\\
&$\llp{121}{131}$ &4.6$\times10^{-5}$  & 1.0 $\times 10^{-2}$
\\
&$\llp{121}{231}$ &4.6$\times10^{-5}$  & 8.8 $\times 10^{-3}$
\\
&$\llp{122}{213}$ &4.6$\times10^{-5}$  & 4.8 $\times 10^{-4}$
\\
&$\llp{132}{313}$ &4.6$\times10^{-5}$  & 1.2 $\times 10^{-3}$
\\
&$\llp{132}{331}$ &4.6$\times10^{-5}$  & 2.6 $\times 10^{-2}$
\\
&$\llp{231}{313}$ &4.6$\times10^{-5}$  & 1.1 $\times 10^{-3}$
\\
&$\llp{231}{331}$ &4.6$\times10^{-5}$  & 2.3 $\times 10^{-2}$
\\
\hline
$B_d \rightarrow e^+\tau^- + e^- \tau^+$& 
$\llp{123}{213}$ & 4.9$\times 10^{-4}$ & 4.8 $\times 10^{-4}$
\\
&$\llp{123}{231}$ &4.9$\times 10^{-4}$  & 8.8 $\times 10^{-3}$
\\
&$\llp{131}{113}$ &4.9$\times 10^{-4}$  & 1.2 $\times 10^{-3}$
\\
&$\llp{131}{131}$ &4.9$\times 10^{-4}$  & 2.6 $\times 10^{-2}$
\\
&$\llp{133}{313}$ &4.9$\times 10^{-4}$  & 1.2 $\times 10^{-5}$
\\
&$\llp{133}{331}$ &4.9$\times 10^{-4}$  & 2.6 $\times 10^{-4}$
\\
&$\llp{231}{213}$ &4.9$\times 10^{-4}$  & 1.1 $\times 10^{-3}$
\\
&$\llp{231}{231}$ &4.9$\times 10^{-4}$  & 2.0 $\times 10^{-2}$
\\
&$\lplp{131}{333}$ & 5.8$\times 10^{-3}$ & 6.8 $\times 10^{-2}$
\\
\hline
$B_d \rightarrow \mu^+\tau^- + \mu^-\tau^+$& 
$\llp{123}{113}$ & 6.0$\times 10^{-4}$ & 4.8 $\times 10^{-4}$
\\
&$\llp{123}{131}$ &6.0$\times 10^{-4}$  & 1.0 $\times 10^{-2}$
\\
&$\llp{132}{113}$ &6.0$\times 10^{-4}$  & 1.2 $\times 10^{-3}$
\\
&$\llp{132}{131}$ &6.0$\times 10^{-4}$  & 2.6 $\times 10^{-2}$
\\
&$\llp{232}{213}$ &6.0$\times 10^{-4}$  & 1.1 $\times 10^{-3}$
\\
&$\llp{232}{231}$ &6.0$\times 10^{-4}$  & 2.0 $\times 10^{-2}$
\\
&$\llp{233}{313}$ &6.0$\times 10^{-4}$  & 1.1 $\times 10^{-3}$
\\
&$\llp{233}{331}$ &6.0$\times 10^{-4}$  & 2.3 $\times 10^{-2}$
\\
&$\lplp{231}{333}$ & 7.4$\times 10^{-3}$ & 5.7 $\times 10^{-2}$
\\
&$\lplp{233}{331}$ &7.4$\times 10^{-3}$ & 1.1 $\times 10^{-1}$
\\
\hline
$B_d \rightarrow \mu^+\mu^-$& 
$\llp{122}{113}$ & 2.4$\times 10^{-5}$ & 4.8 $\times 10^{-4}$
\\
&$\llp{122}{131}$ &2.4$\times 10^{-5}$  & 1.0 $\times 10^{-2}$
\\
&$\llp{232}{313}$ &2.4$\times 10^{-5}$  & 1.1 $\times 10^{-3}$
\\
&$\llp{232}{331}$ &2.4$\times 10^{-5}$  & 2.3 $\times 10^{-2}$
\\
\hline
$B_s \rightarrow \mu^+\mu^-$&
$\llp{122}{123}$ & 5.5$\times 10^{-5}$ & 4.8 $\times 10^{-4}$
\\
&$\llp{122}{132}$ &5.5$\times 10^{-5}$  & 1.6 $\times 10^{-2}$
\\
&$\llp{232}{323}$ &5.5$\times 10^{-5}$  & 1.1 $\times 10^{-3}$
\\
&$\llp{232}{332}$ &5.5$\times 10^{-5}$  & 2.3 $\times 10^{-2}$
\\
\end{tabular}
\end{table}


\begin{references}
\bibitem{ago}
\plb{J. Ellis {\it et al.}}{150}{142}{1985};
\plb{G. G. Ross and J. W. F. Valle}{151}{375}{1985};
\npb{S. Dawson}{261}{297}{1985};
\plb{S. Dimopoulos and L. Hall}{207}{210}{1987}.
\bibitem{hinkae}
\prd{I. Hinchliffe and T. Kaeding}{47}{279}{1993}.
\bibitem{smivi}
\plb{A. Yu. Smirnov and F. Vissani}{380}{317}{1996}.
\bibitem{thopocarl}
\rep{E. H. Thorndike and R. A. Poling}{157}{183}{1987};
\plb{C. E. Carlson, P. Roy and M. Sher}{357}{99}{1995}.
\bibitem{hasu}
\npb{L. J. Hall and M. Suzuki}{231}{419}{1984};
F. Vissani, {\it Supersymmetry '96 Theorectical Perspectives and Experimental
Outlook,} Proceedings of the International Conference, College Park, Maryland,
edited by R. M. Mohapatra and A. Rasin [Nucl. Phys. (Proc. Supp.) {\bf B52A},
94 (1997)].
\bibitem{choroy}
\plb{D. Choudhury and P. Roy}{378}{153}{1996}.
\bibitem{han}
\prd{V. Barger, G. F. Giudice and T. Han}{40}{2987}{1989}.
\bibitem{beta}
\prd{R. N. Mohapatra}{34}{3457}{1986};
\prl{M. Hirsch, H. V. Klapdor--Kleingrothaus and S. G.
Kovalenko}{75}{17}{1995}.
\bibitem{numass}
\npb{R. M. Godbole, P. Roy and X. Tata}{401}{67}{1993}.
\bibitem{agagra}
\prd{K. Agashe, M. Graesser}{54}{4445}{1996}.
\bibitem{dross}
\npb{H. Dreiner and G. G. Ross}{410}{188}{1993}.
\bibitem{bha}
G. Bhattacharyya, preprint IFUP-TH-52/96, hep-ph/9608415.
\bibitem{chahu}
\plb{M. Chaichian and K. Huitu}{384}{157}{1996}.
\bibitem{CLEO}
\prd{R. Ammar {\it et al.}, CLEO Collaboration}{49}{5701}{1994}.
\bibitem{CDF}
CDF Collaboration, Fermilab Conf. 95/201-E(1995).
\bibitem{ali}
A. Ali, hep-ph/9606324.
\end{references}
\end{document}